\documentclass[aps,prb,twocolumn,showpacs]{revtex4}
\usepackage{graphicx}
\begin{document}

\title{Superconducting and normal-state interlayer-exchange-coupling in
La$_{0.67}$Sr$_{0.33}$MnO$_{3}$-YBa$_{2}$Cu$_{3}$O$_{7}$-La$_{0.67}$Sr$_{0.33}$MnO$_{3}$
epitaxial trilayers.}


\author{K. Senapati}
\author{R. C. Budhani}
\email{rcb@iitk.ac.in}
\affiliation{Department of Physics, Indian Institute of Technology
Kanpur, Kanpur - 208016, India}

\date{\today}


\begin{abstract}
The issue of interlayer exchange coupling in magnetic multilayers
with superconducting (SC) spacer is addressed in
La$_{0.67}$Sr$_{0.33}$MnO$_{3}$ (LSMO) - YBa$_{2}$Cu$_{3}$O$_{7}$
(YBCO) - La$_{0.67}$Sr$_{0.33}$MnO$_{3}$ (LSMO) epitaxial
trilayers through resistivity, ac-susceptibility and magnetization
measurements. The ferromagnetic (FM) LSMO layers possessing
in-plane magnetization suppress the critical temperature (T$_{c})$
of the c-axis oriented YBCO thin film spacer. The superconducting
order, however, survives even in very thin layers (thickness
d$_{Y} \sim $ 50 {\AA}, $\sim $ 4 unit cells) at T $<$ 25 K. A
predominantly antiferromagnetic (AF) exchange coupling between the
moments of the LSMO layers at fields $<$ 200 Oe is seen in the
normal as well as the superconducting states of the YBCO spacer.
The exchange energy J$_{1}$ ($\sim $ 0.08 erg/cm$^{2}$ at 150 K
for d$_{Y}$ = 75 {\AA}) grows on cooling down to T$_{c}$, followed
by truncation of this growth on entering the superconducting
state. The coupling energy J$_{1}$ at a fixed temperature drops
exponentially with the thickness of the YBCO layer. The
temperature and d$_{Y}$ dependencies of this primarily
non-oscillatory J$_{1}$ are consistent with the coupling theories
for systems in which transport is controlled by tunneling. The
truncation of the monotonic T dependence of J$_{1}$ below T$_{c}$
suggests inhibition of single electron tunneling across the
CuO$_{2}$ planes as the in-plane gap parameter acquires a non-zero
value.
\end{abstract}


\maketitle


\section{Introduction}
The oscillatory nature of exchange coupling between two
ferromagnetic (FM) layers separated by a metallic but non-magnetic
(NM) spacer as a function of the spacer thickness d$_{n}$ is now
well established in a variety of systems
\cite{ref1,ref2,ref3,ref4,ref5,ref6,ref7}. It is generally agreed
that the coupling is driven by the Rudderman-Kittel-Kasuya-Yoshida
(RKKY)-type exchange through the conduction electrons of the
spacer \cite{ref1}. The period of oscillations predicted by the
theories of exchange coupling is directly related to the extremal
wave vectors connecting opposite sides of the Fermi Surface (FS)
of the spacer material in the direction of the layer growth.
Clearly, the nature of the Fermi surface of the spacer plays a key
role in interlayer exchange. \v{S}ipr and Gy\"{o}rffy \cite{ref8}
first suggested that an experiment in which the Fermi surface
could be changed while keeping all other material parameters the
same, would allow a direct test of the exchange coupling theories
based on extremal wave vectors of the FS. They proposed the use of
a superconducting (SC) spacer in which an isotropic gap opens up
at the FS on cooling below the critical temperature T$_{c}$. The
zero-temperature numerical calculations of \v{S}ipr and
Gy\"{o}rffy \cite{ref8} show that the oscillatory coupling is
strongly damped in the presence of a superconducting gap.
Similarly, the analytical results of de Melo \cite{ref9,ref10}
show that at $\Delta $/T $ >
> $ 1 the coupling decays exponentially as $
{\rm{exp}}\left( {{{{\rm{ - k}}_{{\rm{F}}_{\rm{S}} }
{\rm{d}}_{\rm{s}} \Delta } \mathord{\left/
 {\vphantom {{{\rm{ - k}}_{{\rm{F}}_{\rm{S}} } {\rm{d}}_{\rm{s}} \Delta } {{\rm{E}}_{\rm{F}} }}} \right.
 \kern-\nulldelimiterspace} {{\rm{E}}_{\rm{F}} }}} \right)
$, where k$_{F_{s}}$, d$_{s}$, $\Delta $ and E$_{F}$ are the Fermi
wave-vector, spacer thickness, gap parameter and Fermi energy
respectively. Near T$_{c}$ the large thermally excited
quasiparticle density compensates for the loss of coupling seen at
low temperatures.

Experimental verification of these predictions is, however,
constrained by several materials related factors. First of all,
since the oscillatory coupling is seen only when the spacer
thickness is small ($ \le $ 130 {\AA}), one must ensure that
superconductivity survives in such thin spacers in the presence of
the strong pair-breaking effects of the ferromagnetic boundaries.
Naturally, short coherence length $\xi _{0} $ and high critical
temperature T$_{c}$ of the superconductor, and small exchange
energy of the ferromagnet are the desirable features to see the
effect. In addition, one must also ensure that the interfaces
between the ferromagnetic and superconducting layers are
atomically smooth.

The doped Mott insulators of the perovskite oxide family meet some
of these material specifications. For example,
YBa$_{2}$Cu$_{3}$O$_{7}$ (YBCO) superconductor and
La$_{0.67}$Sr$_{0.33}$MnO$_{3}$ (LSMO) ferromagnet can be grown
epitaxially on top of each other. The cuprate has anisotropic but
short coherence length and a high T$_{c}$, whereas the manganite
with a Curie temperature of $ \approx $ 360 K, has relatively
small exchange energy J ($\sim $ 1 meV) as compared to the J of 3d
transition metal ferromagnets such as Fe and Co which are strong
pair-breakers \cite{ref11}. However, the cuprates also pose
interesting challenges, such as the nodal gap parameters $\Delta $
and anomalous c-axis transport, not present in elemental
superconductors. Reported measurements on high T$_{c}$
superconductor (HTSC)-manganite heterostructures have primarily
focused on the suppression of T$_{c}$ \cite{ref12,ref13,ref14},
spin injection \cite{ref15,ref16,ref17} and the effects on
magnetoresistance \cite{ref18,ref19}. Recently, the aspect of
exchange coupling across HTSC layers has been addressed by
Przyslupski et al. \cite{ref20} using
(La$_{0.67}$Sr$_{0.33}$MnO$_{3})_{n}$/(YBa$_{2}$Cu$_{3}$O$_{7})_{m}$
multilayers where n = 16 unit cells, and m varies from 1 to 8 unit
cells. Measurements of field-cooled (FC) and zero-field-cooled
(ZFC) magnetization loops in these samples reveal exchange biasing
effects, which have been argued to be an indicator of interlayer
exchange coupling. However, this work also attributes the shift of
the FC and ZFC loops to antiferromagnetism in LSMO. Here it needs
to be pointed out that these multilayers have been deposited on
LaAlO$_{3}$ substrates, which introduce large compressive stress
in LSMO and YBCO epitaxial films due to its smaller lattice
parameter ($\sim $ 3.79 {\AA}). In addition, LaAlO$_{3}$ is a
heavily twinned material. Since both these factors are known to
affect magnetic anisotropy of LSMO \cite{ref21} and
superconducting properties of YBCO \cite{ref22}, intrinsic
behavior of FM-SC-FM structure is likely to get masked by such
stress and interface related effects. Further, the interface
related non-intrinsic behavior is likely to get accentuated in
superlattices due to the presence of a large number of interfaces
in such structures.

Here we report the magnetic behavior of LSMO-YBCO-LSMO trilayers,
synthesized on [001] SrTiO$_{3}$ substrates. The lattice parameter
of SrTiO$_{3}$ ($\sim $ 3.91 {\AA}) compares well with the lattice
parameter of L$_{0.67}$Sr$_{0.33}$MnO$_{3}$ ($\sim $ 3.89 {\AA})
and the average ab-plane lattice spacing of
YBa$_{2}$Cu$_{3}$O$_{7}$ ($\sim $ 3.85 {\AA}). The scope for a
stress-free layer-by-layer growth has been improved further
through special chemical treatment of the substrate \cite{ref23}.
We first describe the magnetic behavior of plane LSMO films of
various thickness. High quality epitaxial layers of LSMO showing a
soft magnetic character and in-plane anisotropy were integrated
with YBCO in a trilayer form and the superconducting critical
temperature of such structures was measured. The suppression in
T$_{c}$ has been attributed to the pair breaking effects at the
FM-SC boundary. Finally, the issue of interlayer exchange coupling
has been addressed through measurements of ZFC in-plane
magnetization loops over a broad range of temperatures. These
measurements reveal a long range antiferromagnetic coupling
between LSMO layers decaying exponentially with the thickness of
the YBCO spacer.

\section{experimental}

Thin films of LSMO, and trilayers of LSMO-YBCO-LSMO and
PrBa$_{2}$Cu$_{3}$O$_{7} $ - YBa$_{2}$Cu$_{3}$O$_{7} $ -
PrBa$_{2}$Cu$_{3}$O$_{7}$ (PBCO-YBCO-PBCO) were deposited on
chemically polished [001] oriented SrTiO$_{3}$ substrates. A
multitarget pulsed laser deposition technique based on KrF excimer
laser ($\lambda $= 248 nm) was used to deposit the films and
trilayers at 800 $^{o}$C and 200 mTorr O$_{2}$ partial pressure
\cite{ref24}. Since the growth conditions for all three oxides
were identical, the trilayers were deposited sequentially without
changing the process parameters. A slow deposition rate ($\sim $ 1
{\AA}/sec) established through several calibration runs, was used
to realize a layer-by-layer growth of LSMO, PBCO, and YBCO. While
for the plane LSMO films we have studied the changes in transport
and magnetic properties as a function of thickness from 50 {\AA}
to 1100 {\AA}, the thickness of each LSMO layer in LSMO-YBCO-LSMO
trilayers was fixed at 300 {\AA}, and the thickness of the cuprate
was varied from 50 {\AA} to 300 {\AA}. For the PBCO-YBCO-PBCO
trilayers, a constant PBCO layer thickness of 100 {\AA} was used.
The crystallographic structure of the films was characterized with
$\theta - 2\theta $ X-ray diffraction. The SC and FM critical
temperatures of the films were established through resistivity
$\rho (T)$, ac-susceptibility $\chi (T)$ and magnetization M(T)
measurements. We have used a home-built micro-Hall-probe based
ac-susceptometer \cite{ref25} for detailed measurements of vortex
dynamics in these films \cite{ref24}. The measurements of
resistivity in the temperature range of 2 K to 370 K were carried
out in the standard four-probe geometry. A superconducting quantum
interference device (SQUID) based magnetometer (MPMS-XL5) operated
in the RSO mode for higher sensitivity was used for detailed
measurements of zero-field-cooled and field-cooled magnetization
and M-H loops. The magnetic field in these measurements was in the
plane of the film, aligned along one of the principal axes ([100]
or [010]). Measurements were also performed with the field in the
[110] direction to check for the in-plane anisotropy of
magnetization.

 \begin{figure}
 \includegraphics [width=7cm]{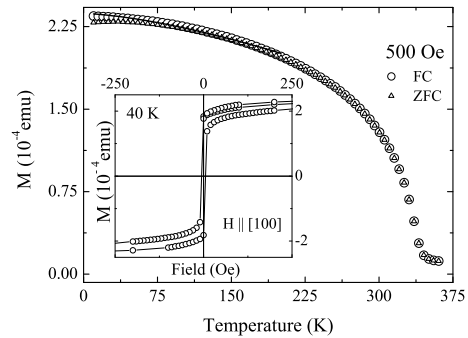}%
 \caption{\label{fig 1}Zero-field-cooled ($\Delta )$ and field-cooled (${\rm O})$
magnetization of a 600 {\AA} thick film of
La$_{0.67}$Sr$_{0.33}$MnO$_{3}$ measured with 500 Oe in-plane
field directed along the [100] axis. The solid line is a fit to
the field-cooled curve using the Bloch law (Eq. 1) for the decay
of magnetization at T $<<$ T$_{c}$. Inset: Zero-field-cooled
hysteresis loop of the same film measured at 40 K in the same
configuration as the measurement of $M\left( {T} \right)$.}
 \end{figure}

\section{results}
\subsection{Magnetic ordering in thin La$_{0.67}$Sr$_{0.33}$MnO$_{3}$ films:}

 \begin{figure}
 \includegraphics[width=6cm]{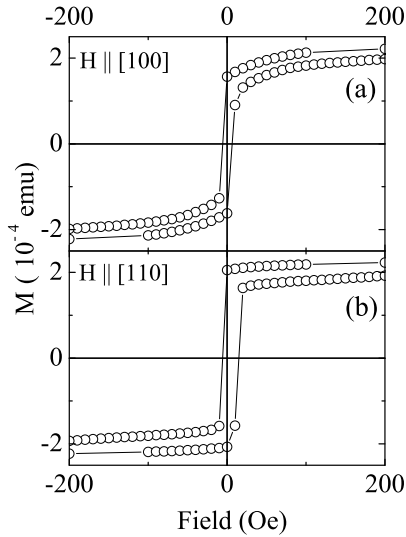}%
 \caption{\label{fig2}Zero-field-cooled magnetization loops at 100 K measured with
in-plane applied field along the [100] (panel `a') and [110]
(panel `b') directions. In both cases the data are corrected for a
small diamagnetic contribution from the STO substrate.}
 \end{figure}

Figure 1 shows the ZFC and FC magnetization of a 600 {\AA} thick
LSMO film measured at 500 Oe. The onset of spontaneous
magnetization at $\sim $ 350 K on cooling marks the Curie
temperature of the sample. The ZFC and FC branches of
magnetization in granular and multi-domain magnetic films of large
coercivity show a pronounced bifurcation at lower temperatures. In
the present case, however, the two branches nearly superimpose
down to the lowest temperature. This feature indicates the growth
of a defect-free magnetic film of low coercivity. We analyze the
temperature dependence of the FC magnetization in the framework of
the Bloch theory for decay of magnetization due to excitation of
spin waves \cite{ref26}. The drop in saturation magnetization is
predicted to be of the form;

\begin {equation}
{{M_{s} \left( {T} \right)} \mathord{\left/ {\vphantom {{M_{s}
\left( {T} \right)} {M_{s} \left( {0} \right) = 1 - AT^{{{3}
\mathord{\left/ {\vphantom {{3} {2}}} \right.
\kern-\nulldelimiterspace} {2}}}}}} \right.
\kern-\nulldelimiterspace} {M_{s} \left( {0} \right) = 1 -
AT^{{{3} \mathord{\left/ {\vphantom {{3} {2}}} \right.
\kern-\nulldelimiterspace} {2}}}}}
\end{equation}

\noindent Here M$_{s}$(0) is the saturation magnetization at T =
0, and the coefficient A is expressed as$\left( {{{C}
\mathord{\left/ {\vphantom {{C} {S}}} \right.
\kern-\nulldelimiterspace} {S}}} \right)\left( {{{k_{B}}
\mathord{\left/ {\vphantom {{k_{B}} {2JS}}} \right.
\kern-\nulldelimiterspace} {2JS}}} \right)^{{{3} \mathord{\left/
{\vphantom {{3} {2}}} \right. \kern-\nulldelimiterspace} {2}}}$,
where C = 0.059 for a simple cubic magnetic lattice, S the total
spin and J the exchange integral which is given by the formula
${{k_{B} T_{c}} \mathord{\left/ {\vphantom {{k_{B} T_{c}}  {J}}}
\right. \kern-\nulldelimiterspace} {J}} = \left( {{{5}
\mathord{\left/ {\vphantom {{5} {96}}} \right.
\kern-\nulldelimiterspace} {96}}} \right)\left( {Z - 1}
\right){\left[ {11S\left( {S + 1} \right) - 1} \right]}$of
Rushbrooke and Wood \cite{ref27}. An excellent fit to the
magnetization at T $<<$ T$_{Curie}$ is seen with a dependence of
the type $1 - AT^{{{3} \mathord{\left/ {\vphantom {{3} {2}}}
\right. \kern-\nulldelimiterspace} {2}}}$, when the average spin S
(= 1.835) per Mn site is used. The exchange energy deduced from
the fits is $\sim $ 2 meV, while the exchange energy of the strong
ferromagnets like Fe is $\sim $ 5.5 meV deduced from a Bloch
constant of $\sim $ 3.6x10$^{ - 6}$ deg$^{ - 3 / 2} $\cite{ref26}.

\begin{figure}
 \includegraphics[width=7cm]{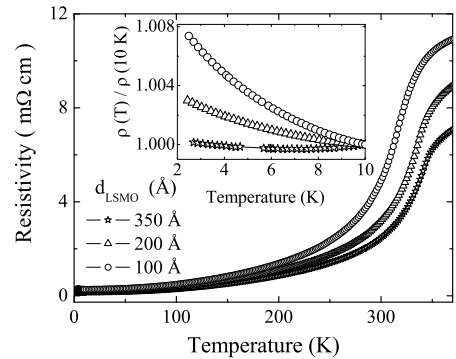}%
 \caption{\label{fig3}Resistivity ($\rho \left( {T} \right))$ of LSMO films deposited on
STO in the temperature range of 2 K - 370 K. Thickness of the
films varies from 100 {\AA} to 350 {\AA}. Inset shows a magnified
view of the low temperature section of the $\rho \left( {T}
\right)$ curves. These data have been normalized with respect to
the resistivity at 10 K in order to emphasize the upturn in the
resistivity of the thinnest films.}
 \end{figure}

In order to check for a preferred in-plane axis of magnetization,
we have measured the hysteresis loops with the external field
aligned along the [100] and [110] directions of the [001] oriented
films. Results of this measurement are shown in Fig. 2(a, b). A
perfect hysteresis loop with the remanant magnetization (M$_{r})$
equal to M$_{s}$ is seen when the field is along [110]. Whereas in
the case of H $\vert \vert $ [100] (Fig. 2a), $M_{r} = {{M_{s}}
\mathord{\left/ {\vphantom {{M_{s}}  {\sqrt {2}}} } \right.
\kern-\nulldelimiterspace} {\sqrt {2}}} $. This observation
clearly indicates that [110] is the easy axis of magnetization and
[100] is the hard axis. However, the small value of the switching
field suggests that the energy barrier for rotation of
magnetization is not large. This result is consistent with the
earlier measurements of magnetization loops in films of LSMO
deposited on STO substrates \cite{ref28,ref29}. The square
hysteresis loops seen in the figure further suggest that these
films are magnetically quite soft. The behavior of magnetization
in LSMO films deposited on LAO is quite different. The preferred
direction of magnetization is perpendicular to the film plane in
this case \cite{ref21,ref30,ref31}.

Unlike the other double exchange manganites such as
La$_{0.67}$Ca$_{0.33}$MnO$_{3}$, the resistivity of LSMO with 30
to 40 {\%} Sr is metallic in the paramagnetic state \cite{ref32}.
This metallic conduction is seen in our films as well. The
resistivity of these films at room temperature is low ($\sim $ 2
m$\Omega $ cm), and remains metallic down to 2 K. Fig. 3 displays
the zero-field resistivity of LSMO films spanning over a thickness
range of 100 {\AA} to 350 {\AA} in the temperature window of 2 K -
370 K. The paramagnetic metallic phase above T$_{Curie}$ which
transits to a ferromagnetic metallic phase at T $<$ T$_{Curie}$,
is clearly identifiable for all films. T$_{Curie}$ acquires the
near bulk value ($\sim $ 350 K) for films thicker than 200 {\AA},
while thinner films show a slight drop in the Curie temperature.
The resistivity at the lowest temperature normalized with respect
to its value at 10 K, is shown in the inset of Fig. 3. A small
upturn in resistivity, which can be attributed to weak
localization and electron-electron interaction effects in 2D, is
observed only for the thinnest films ($ \le $ 200 {\AA}). These
features indicate the growth of a high quality thin film of LSMO.
The magnetic and electrical characteristics of LSMO dominate the
behavior of $\rho (T)$ and $M(T)$ in LSMO-YBCO-LSMO trilayers at T
$ > $ 100 K as described in the following section.

\subsection{Superconductivity in LSMO - YBCO - LSMO trilayers:}

 \begin{figure}
 \includegraphics[width=7cm]{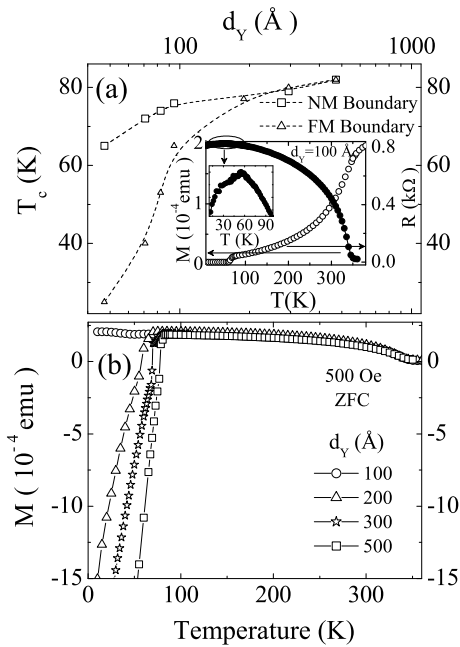}%
 \caption{\label{fig4}Panel (a) : T$_{c}$ (open symbols) plotted as a function of
d$_{Y}$ in LSMO-YBCO-LSMO and PBCO-YBCO-PBCO trilayers. Inset
shows $R\left( {T} \right)$ and zero-field-cooled $M\left( {T}
\right)$ of a LSMO-YBCO-LSMO sample with d$_{Y}$ = 100 {\AA}. An
enlarged view of $M\left( {T} \right)$ near T$_{c}$ is also shown.
Panel (b): Temperature dependence of the zero-field-cooled
magnetization of LSMO-YBCO-LSMO trilayers with 100, 200, 300 and
500 {\AA} thick SC layers. The measurement field of 500 Oe was
applied along the [100] direction in the plane of the films.}
 \end{figure}

In the inset of Fig. 4(a) we plot the magnetization and
resistivity of a LSMO-YBCO-LSMO trilayer with YBCO layer thickness
(d$_{Y})$ of 100 {\AA}. At T $ \le $ 360 K a metallic behavior is
evident in the resistivity plot. This becomes pronounced at T $<$
T$_{Curie}$. At lower temperatures however, the resistance of the
sample drops to zero as the current path is shorted by the
superconducting YBCO layer. Correspondingly, there is a non-zero
diamagnetic contribution to magnetization due to the Meissner
effect. In trilayers with thicker YBCO film, the magnetization
actually crosses the zero-line and becomes negative. This is seen
in Fig. 4(b) where we have plotted the ZFC magnetization of some
trilayers with different YBCO thickness. The superconducting
transition temperature (T$_{c})$ in these heterostructures is a
strong function of YBCO layer thickness. In Fig. 4(a) we show the
variation of T$_{c}$ as a function of d$_{Y}$ in LSMO-YBCO-LSMO
trilayers. In order to estimate the effects of magnetic boundaries
on T$_{c}$, we have also measured the superconducting transition
temperature of PBCO-YBCO-PBCO trilayers. Results of these
measurements are also shown in Fig. 4(a). For this non-magnetic
system, the T$_c$ drops as the thickness of the YBCO layer
(d$_{Y}$) is reduced. The variation of T$_{c}$ with d$_{Y}$ in
PBCO-YBCO-PBCO multilayers has been studied extensively by several
groups, and various reasons have been given for the drop
\cite{ref33,ref34,ref35}. These include interfacial stress, a drop
in c-axis coupling of the condensate as the number of CuO$_{2}$
planes is reduced etc. However, the effect of uniaxial stress
applied along the a-axis and b-axis of the YBCO crystal on its
T$_{c}$ is nearly equal and opposite \cite{ref36}. This result
rules out any direct effect of the lattice mismatch induced stress
on T$_{c}$. However, Varela et al \cite{ref34} have shown that the
overall stress  pattern also gives rise to very significant and
non-uniform changes within YBCO unit cell, which may reduce the
hole concentration in the CuO$_2$ planes located close to the
interfaces. We may write this interface driven reduction in T$_c$
as $\Delta$T$_c$(d$_Y$)$_{interface}$. Since the lattice
parameters of La$_{0.67}$Sr$_{0.33}$MnO$_{3}$ and
PrBa$_{2}$Cu$_{3}$O$_{7}$ are identical within 0.5\%, we assume
the effect of the interface on T$_c$ to be similar for YBCO films
sandwiched between the LSMO layers. The LSMO layers, however, also
affect the T$_c$ through pair-breaking. We can therefore argue
that the larger drop in T$_c$ of LSMO-YBCO-LSMO trilayers of a
given d$_Y$ as compared to the T$_c$ of PBCO-YBCO-PBCO of the same
d$_Y$ is due to the magnetic pair-breaking effects. A rigorous
treatment of pair breaking effects of a ferromagnetic film
deposited on top of a superconductor requires solution of the
Usadel equation for different degree of interface transparency for
Cooper pair tunneling\cite{refRadovic,refLazar}.

\subsection{Magnetic coupling in LSMO - YBCO - LSMO trilayers:}

 \begin{figure}
 \includegraphics[width=8cm]{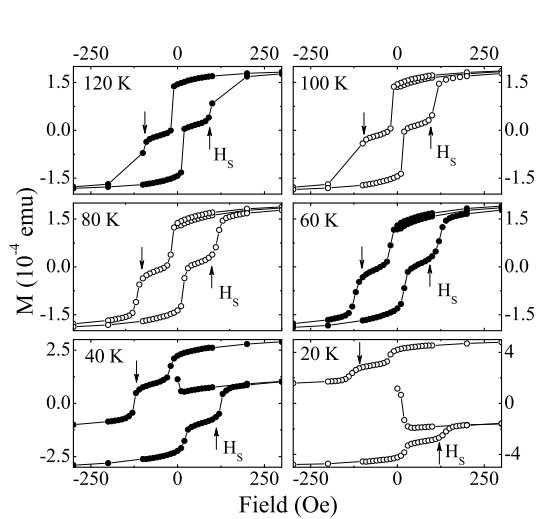}%
 \caption{\label{fig5}Low-field section of isothermal hysteresis loops of a
LSMO-YBCO-LSMO trilayer with 100 {\AA} YBCO interlayer, measured
at 20, 40, 60, 80, 100, and 120 K. All measurements were performed
on zero-field-cooled samples and with in-plane applied field along
the [100] direction. The switching field has been marked as
H$_{s}$ (see text for details).}
 \end{figure}

Having established the existence of ferromagnetic and
superconducting orders in these trilayers, we now discuss the
behavior of interlayer magnetic coupling between the LSMO layers
separated by YBCO below and above the T$_{c}$. Fig. 5 shows a
series of M-H loops of an LSMO-YBCO-LSMO trilayer with d$_{Y}$ =
100 {\AA} taken at various temperatures with the external magnetic
field aligned along [100] direction. In the loops measured at T $
> $ 100 K, one can easily identify a critical field $\vert
$H$_{s}$$\vert $ up to which the magnetic moment of the trilayer
remains close to zero, and then quickly achieves the saturation
value once the field $\vert $H$\vert $ exceeds $\vert
$H$_{s}$$\vert $. The magnetization of the sample below the
superconducting transition drops rapidly at low fields because of
the diamagnetic signal from the YBCO layer. The reverse branch of
the hysteresis loops shows a large irreversibility due to pinning
of the magnetic flux. However, the ferromagnetic component of the
magnetization is also found to persist in the superconducting
state. A careful look at the magnetizing branches of Panels c, d
and e clearly reveals the characteristic field H$_{s}$ below which
the magnetization remains nearly constant. A similar behavior of
the hysteresis, both above and below T$_{c}$,$_{} $is seen in
trilayers with a different YBCO layer thickness. The M-H loops of
some samples at 100 K are shown in Fig. 6.

A straightforward explanation for the existence of H$_{s}$ can be
given by invoking antiferromagnetic exchange coupling between the
LSMO layers separated by normal and superconducting YBCO. Earlier
measurements of magnetization in superlattices of ferromagnetic
manganites and non-magnetic but metallic LaNiO$_{3}$ have revealed
indirect coupling mediated by the conduction electrons of the
spacer \cite{ref6,ref7}. This coupling is oscillatory with the
thickness of the spacer. However, before proposing such
fundamental mechanism, we must rule out some rather mundane
processes, which could lead to a similar effect. First of all, the
interfaces of perovskite oxides based multilayers have an inherent
stereochemical disorder even when they are atomically sharp
\cite{ref37}. This disorder is caused by a change in the nearest
neighbor environment of the magnetically active ions, and can lead
to random pinning of their moment. While a gradual depinning of
these moments with the increasing field would lead to deviations
from a square hysteresis loop typical of a soft magnet such as
LSMO, it is not likely to result in the M-H loops seen in Figs. 5
and 6. One may also argue that the uncompensated copper spins at
the interface help stabilize this random state. In order to rule
out these possibilities, we have measured the M-H loops of
LSMO-PBCO-LSMO trilayers. In Fig. 7(a, b and c) we compare the
magnetization curves of a 600 {\AA} thick single layer LSMO film,
a LSMO-PBCO-LSMO trilayer and a LSMO-YBCO-LSMO trilayer. In all
three cases the measuring field was along the [100] direction.
Fig. 7(d) shows the M-H loop of the LSMO-YBCO-LSMO trilayer
measured with H $\vert \vert $ [110] for comparison. It is evident

 \begin{figure}
 \includegraphics[width=8cm]{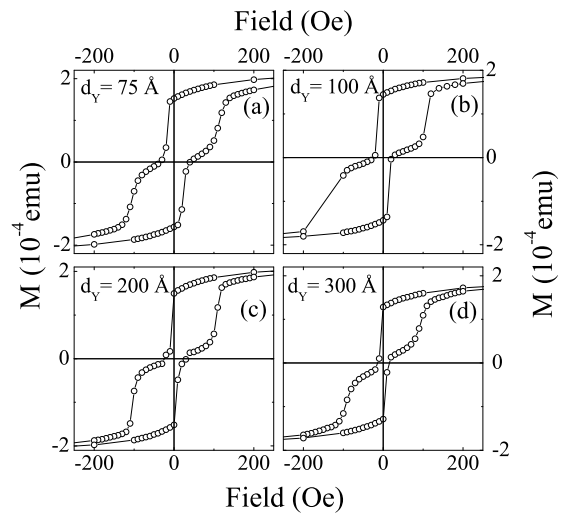}%
 \caption{\label{fig6}Panels `a', `b', `c' and `d', display the results of isothermal
magnetization measurements at 100 K for trilayers with 75, 100,
200 and 300 {\AA} thick YBCO interlayer, respectively. The
measurement field was directed along the [100] direction in the
plane of the trilayers. The low field region is magnified in order
to emphasize the antiferromagnetic coupling between the LSMO
layers below 200 Oe.}
 \end{figure}

from these data that the hysteresis with the characteristic
magnetizing field H$_{s}$ is seen only in the case of
LSMO-YBCO-LSMO trilayers. This observation rules out the role of
uncompensated copper spins, as these factors are present in
LSMO-PBCO-LSMO as well. Some signatures of the type of M-H curve
seen in Figs. 5 and 6, have also been observed by Przyslupski et
al. \cite{ref20} in La$_{0.67}$Sr$_{0.33}$MnO$_{3}$ --
YBa$_{2}$Cu$_{3}$O$_{7}$ superlattices with thin ($\sim $ 60
{\AA}) LSMO layers. They have presented a scenario where migration
of holes from the YBCO into the LSMO converts a few unit cells of
the latter into an antiferromagnet. This AF ordered layer pins the
magnetic moment of the remaining ferromagnetic portion. Since the
LSMO layer is on both sides of the YBCO, this effect should lead
to two pinned magnetization vectors whose relative orientation can
be anywhere from 0 to 180$^{o}$. However, the observation of a net
zero magnetization at H $<$ $\vert $H$_{s} \vert $ demands that
this angle is 180$^{o}$. This is possible only when there is an
exchange coupling across the YBCO. The magnetic behavior of
ferromagnetic and antiferromagnetic LSMO couples has been studied
in detail by Izumi et al. \cite{ref38}, for the case of
La$_{0.6}$Sr$_{0.4}$MnO$_{3}$ / La$_{0.45}$Sr$_{0.55}$MnO$_{3}$
superlattices, where the latter compound is a metallic A-type
antiferromagnet. The Mn spins in alternate layers of this compound
are coupled antiferromagnetically with their orientation in the
[001] plane. Izumi et al. \cite{ref38} note that the magnetization
of the ferromagnetic layers is perpendicular to the magnetic easy
axis of the antiferromagnetic layer, thus ruling out exchange
biasing. While the measurement of far infrared conductivity
$\sigma _{1} \left( {\omega}  \right)$ of YBCO-LSMO multilayers by
Holden et al. \cite{ref39} do show a strong suppression of the
free-carrier contribution to $\sigma _{1} \left( {\omega}
\right)$, migration of holes is only one of the possible
mechanisms proposed by them for the suppression. Further, keeping
in view the result of Izumi et al. \cite{ref38}, a possible hole
transfer does not \textit{a priory} imply exchange biasing. In
order to check if there is any exchange biasing effect of the
interfacial LSMO in our trilayers, we have plotted a minor loop
for a sample with d$_{Y}$ = 100 {\AA}. Starting from saturation
magnetization in the forward direction, the field was decreased to
a value $\vert $H$\vert < \vert $H$_{s}$$\vert $ in the negative
direction and then increased again. In the presence of exchange
biasing, the minor loop obtained in this way should be shifted
along the field axis by an amount equal to the biasing field.
However, the minor loop in Fig. 8 shows no shift within an
accuracy of 5 Oe.

\begin{figure}
 \includegraphics[width=9cm]{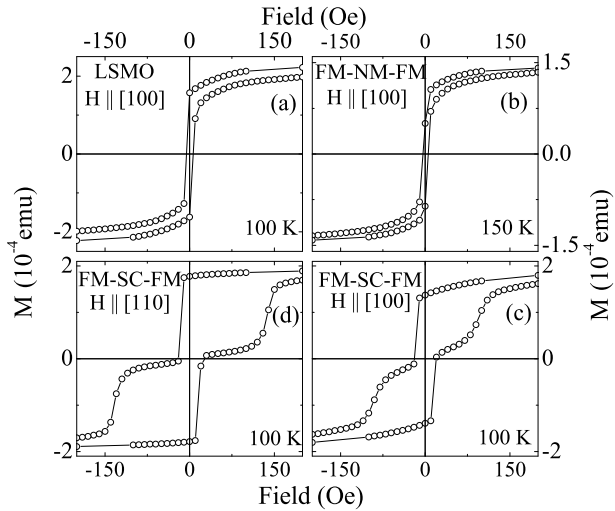}%
 \caption{\label{fig7}A comparative view of the magnetization behavior of a 600 {\AA}
LSMO film (panel `a'), LSMO-PBCO-LSMO trilayer with 300 {\AA} PBCO
(panel `b') and LSMO-YBCO-LSMO trilayer with 300 {\AA} YBCO (panel
`c'). The external magnetic field was in the plane of the film and
directed along the [100] axis. Panel `d' shows the magnetization
while the field was applied along the [110] direction for the
LSMO-YBCO-LSMO trilayer. All the curves were corrected for a small
diamagnetic response of the substrate.}
 \end{figure}

 \begin{figure}
 \includegraphics[width=7cm]{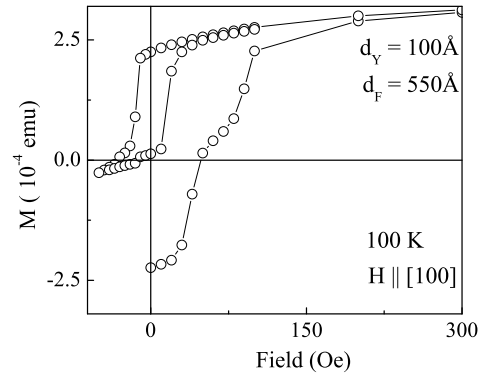}%
 \caption{\label{fig8}The minor hysteresis loop (see text for details) of a trilayer
with 100 {\AA} YBCO layer sandwiched between 550 {\AA} LSMO layers
on both sides, is shown superposed on the main magnetization
curve. The measurement was performed at 100 K after cooling the
sample in zero-field.}
 \end{figure}

Noting that the hysteresis seen in Figs. 6 and 7 is a signature of
antiferromagnetic coupling between the LSMO layer magnetizations,
we now proceed to estimate the exchange energy and its temperature
dependence. The free energy expression for two magnetic layers of
equal thickness coupled by bilinear coupling can be written as
\cite{ref40};

\begin{equation}
F = F_{c} + F_{a} - \overrightarrow {H} .(\overrightarrow {M} _{1}
+ \overrightarrow {M} _{2} )t
\end{equation}

\noindent
where $\overrightarrow {M} _{_{1}}  $ and
$\overrightarrow {M} _{_{2}}  $ are the magnetizations of the top
and bottom LSMO layers, F$_{c}$ is the coupling energy per unit
area, and t the thickness of one LSMO layer. The anisotropy energy
F$_{a}$ derives contributions from the magnetocrystalline
anisotropy as well as in-plane uniaxial anisotropy of the film.
Under the assumption of a bilinear coupling, F$_{c}$ can be
written as;

\begin{equation}
F_{c} = - J_{1} (\widehat{M}_{1}  .\widehat{M}_{2}  )
\end{equation}

\noindent Here $\widehat{M}_{1}  $ and $\widehat{M}_{2} $ are unit
magnetization vectors, and J$_{1} <$ 0 corresponds to
antiferromagnetic coupling between the FM layers. The equilibrium
orientation of $\overrightarrow {M} _{_{1}}  $and $\overrightarrow
{M} _{_{2}}  $ are found by minimization of the free energy with
respect to variations in the orientations of these two vectors. In
a special case, when the interlayer exchange $J_{1}
(\widehat{M}_{1}  .\widehat{M}_{2} )$ is $<$ F$_{a}$, the
magnetization increases slowly in small field and then at a
critical value of the field jumps to the saturation M$_{s}$. The
switching field H$_{s}$ in this case for magnetic layers of equal
thickness (t), and magnetization density (M$_{s}$) is written as;

\begin{equation}
H_{s} = - \left( {J_{1} / M_{s} t} \right)
\end{equation}

\noindent The behavior of magnetization seen in Figs (5, 6, and 7)
corresponds to this situation. We have made an estimate of J$_{1}$
from the measured H$_{s}$ and magnetization density M$_{s}$ using
Eq. 4. Fig. 9 shows the variation of J$_{1}$ as a function of
temperature for trilayers of different YBCO layer thickness. In
the figure we note that the coupling energy at T $ > $ T$_{c}$ is
small, non-oscillatory and decreases exponentially with the
thickness of the superconductor. In all cases however, J$_{1}$
increases monotonically as the temperature is lowered to T$_{c}$.
Below this temperature a truncation of the monotonic growth of
J$_{1}$ is evident in all samples.

The temperature dependence of the interlayer exchange coupling in
metallic multilayers has been worked out theoretically
\cite{ref41, refBruno}. Following Bruno \cite{refBruno}, the
amplitude of the linear exchange coupling coefficient J$_{1}$
increases with the decreasing temperature in the following manner,

\begin{equation}
J_1 \left( T \right) = J_1 \left( 0 \right)\left( {\frac{{{T
\mathord{\left/
 {\vphantom {T {T_0 }}} \right.
 \kern-\nulldelimiterspace} {T_0 }}}}{{\sinh {T \mathord{\left/
 {\vphantom {T {T_0 }}} \right.
 \kern-\nulldelimiterspace} {T_0 }}}}} \right)
\end{equation}

\noindent where the characteristic temperature T$_{0}$ depends on
Fermi wave-vector k$_{F}$ and spacer thickness d$_{n}$ through the
relation $T_{0} = {{\hbar k_{F}}  \mathord{\left/ {\vphantom
{{\hbar k_{F}}  {2\pi k_{B} d_{n} m}}} \right.
\kern-\nulldelimiterspace} {2\pi k_{B} d_{n} m}}$, where m is the
free electron mass and $\hbar $ and k$_{B}$ are Planck and
Boltzmann constants, respectively. The calculations of Edwards et
al. \cite{ref41} also yield a similar temperature dependence of
J$_{1}$. Since the transport in the present case is along the
c-axis of the YBCO, the relevant Fermi wave-vector is $k_{F_{Z}} =
{{\pi} \mathord{\left/ {\vphantom {{\pi} {2c}}} \right.
\kern-\nulldelimiterspace} {2c}}$, where c is the c-axis lattice
parameter ($\sim $ 12 {\AA}) \cite{ref10}. We have fitted the
temperature dependence of J$_{1}$ shown in Fig. 9 to Eq. 5.
However, the average value of $k_{F_{Z}}  $obtained from these
fits is larger by a factor of $\sim $ 4 compared to the $k_{F_{Z}}
$expected for YBCO \cite{ref10}. In Fig. 9 we show a theoretical
curve for J$_{1}$(T) generated using Eq. 5 with $k_{F_Z }  = {\pi
\mathord{\left/
 {\vphantom {\pi  {2c}}} \right.
 \kern-\nulldelimiterspace} {2c}}$, d$_{Y}$=100 {\AA} and J$_{1}$(T)
such that the experimental and calculated values of J$_{1}$ at 120
K are the same. The calculated J$_{1}$(T) shows a steep increase
at the lower temperatures where the experimental data reach
saturation. This truncation of the theoretically expected growth
of J$_{1}$ below T$_{c}$ is suggestive of a superconducting gap.

\section{discussion}

 \begin{figure}
 \includegraphics[width=8cm]{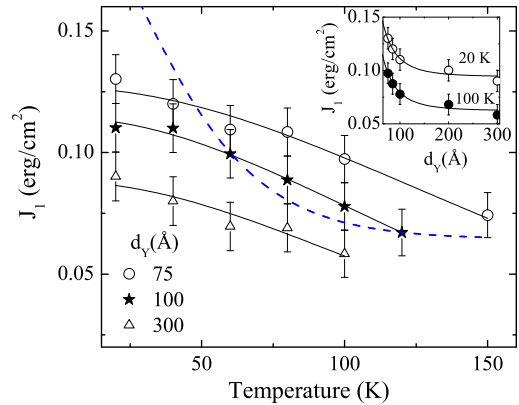}%
 \caption{\label{fig9}The antiferromagnetic coupling energy
(J$_{1})$ calculated using the relation $ \vert J_{1}\vert  =
H_{s} M_{s} t $ (Eq. 4), is shown as a function of temperature for
three sandwich structures with YBCO thickness of 75 {\AA} (O), 100
{\AA} $(\star)$, and 300 {\AA} ($\Delta )$ and a constant LSMO
thickness of 300 {\AA}. The solid lines are fits to the equation $
J_{1} \sim (T/T_{0})/\text{sinh}(T/T_{0})$ (Ref. 44). Dashed line
is a theoretically generated curve showing how J$_1$ should grow
with respect to the value at 120 K if k$_{F_{Z}}$ is taken to be
$\pi /2c$, where c is the c-axis lattice parameter (see text for
details). Inset: The dependence of J$_{1}$ on the thickness of the
YBCO spacer is plotted at 20 K and 100 K. A characteristic decay
length of $\sim $ 150 {\AA} was obtained by fitting these data to
a first order exponentially decaying function (shown as the solid
lines).}
 \end{figure}

Although the physics of magnetic coupling across a superconducting
spacer of anisotropic order parameter is an enormously complicated
problem to analyze, the following arguments can be made on the
basis of the data shown in Fig. 5 through Fig. 9. We first
consider the case when YBCO is in the normal state. The coupling
in this situation is mediated by the transport of carriers
perpendicular to CuO$_{2}$ planes in these c-axis-oriented films.
While the resistivity of YBCO along the c-axis shows a variety of
behaviors depending on doping concentration and defect structure,
for optimally doped YBCO it is metallic, but larger by a factor of
$\sim $ 50 compared to the in-plane resistivity
\cite{refSLcooper}. The c-axis transport in optimally doped and
overdoped  YBCO involves blocking of coherent interplanar
tunneling by the in-plane scattering. This leads to $\rho_c$
$\alpha$ $\rho_{ab}$\cite{refNkumar,refMSkumar}. In underdoped
systems diffusive tunneling dominates the transport, leading  to a
semiconductor like resistivity \cite{refTurlakov}. The
non-oscillatory and predominantly antiferromagnetic IEC seen here
is analogous to the behavior of exchange coupling in Fe-FeSi-Fe
\cite{refdeVries} and Fe-Si-Fe \cite{refStrijkers}
heterostructures. The IEC in this case is strongly
antiferromagnetic (J$_{1}$ $\sim $ 2 erg/cm$^{2})$ for a thin
spacer, and decays exponentially with the increasing spacer
thickness. Furthermore, the exchange energy J$_{1}$ shows a
monotonic drop with the increasing temperature, a behavior similar
to the data shown in Fig. 9. A bias towards antiferromagnetic IEC
has been predicted theoretically as well. Shi, Levy and Fry
\cite{refShi} have shown that this bias for AF-coupling arises
from a competition between the RKKY--like exchange and
superexchange, and a non-cancellation of the non-oscillatory parts
of these two contributions. An AF coupling, which decays
exponentially with the spacer thickness, has been predicted by
Slonczewski \cite{ref49} and Bruno \cite{refBruno} using an
electron tunneling picture. The theory \cite{refBruno} predicts a
d$_{n}$ dependence of the type $J_{1} \sim \left( {{{1}
\mathord{\left/ {\vphantom {{1} {d_{n}^{2}}} } \right.
\kern-\nulldelimiterspace} {d_{n}^{2}}} } \right)\exp \left( {{{ -
d_{n}}  \mathord{\left/ {\vphantom {{ - d_{n}} {\lambda}} }
\right. \kern-\nulldelimiterspace} {\lambda}} } \right)$. The
calculated value of the coupling energy J$_{1}$ for our trilayers
is plotted in the inset of Fig. 9 as a function of the spacer
layer thickness. We have fitted these data to a first-order
exponential decay of the type given by Bruno et al.
\cite{refBruno}. Result of this fitting is shown as solid-lines in
the inset. The characteristic decay length $\lambda $ inferred
from the fit is $\sim $ 150 {\AA}. Since the c-axis transport in
YBCO is controlled by a delicate balance between single electron
tunneling and intralayer electron - electron scattering processes
\cite{refNkumar,refMSkumar,refTurlakov}, a tunneling picture for
IEC is applicable, albeit with the caveat that it is unlike the
tunneling through a semiconducting barrier where thermally induced
carriers can enhance IEC at higher temperatures \cite{refBruno}.
The IEC in this case is expected to decay with temperature as the
c-axis resistivity shows a linear temperature dependence.

The truncation of the monotonic growth of the exchange coupling
energy when the YBCO layer becomes superconducting (as seen in
Fig. 9) is in agreement with the predictions of \v{S}ipr and
Gy\"{o}rffy \cite{ref8}, and of de Melo \cite{ref9}. However, the
extent of the drop in the coupling energy in the T = 0 limit
depends on the strength of the superconducting gap parameter. For
a weak ferromagnet and an isotropic superconductor, de Melo
\cite{ref9} has derived an analytic expression for the effective
coupling Hamiltonian;

\begin{equation}
H_{eff} \sim {\frac{{\cos \left( {2k_{F} d_{s}} \right)}}{{\left(
{2k_{F} d_{s}}  \right)^{2}}}}\exp \left( { - {\frac{{k_{F} d_{s}
\Delta}} {{E_{F} }}}} \right)
\end{equation}
\noindent where, k$_{F}$ and d$_{s}$ are the Fermi momentum and
thickness of the S-layer respectively. $\Delta $ is the
superconducting gap and E$_{F}$ is the Fermi energy. This
expression shows that the superconducting order does not actually
contribute to the oscillating part of the interaction, it only
induces a relative decrease in the strength of interaction as
compared to the interaction for a normal metallic spacer. However,
the low temperature calculations in Refs. 8 and 9 are valid only
for an isotropic gap parameter. de Melo has recently considered
the case of IEC through a d-wave superconductor whose order
parameter lies in the plane of the multilayer \cite{ref10}. The
main contribution to coupling in this case comes from the wave
vectors connecting points at the Fermi surface along the [001]
direction and for $\overrightarrow {k} _{\vert \vert}  $ lying
along the direction of nodes. A distinct suppression, although not
as much as in the case of a fully gapped system, has been seen in
the superconducting state.

\section{Conclusions}
In summary, we have studied the magnetic and superconducting
states of epitaxial thin film heterostructures consisting of two
La$_{0.67}$Sr$_{0.33}$MnO$_{3} $layers separated by a layer of
YBa$_{2}$Cu$_{3}$O$_{7}$, whose c-axis is perpendicular to the
plane of the heterostructure. We see a distinct influence of the
ferromagnetic boundaries on the T$_{c}$ of the YBCO layer. This is
attributed to the pair-breaking phenomena near the F-S interface.
The hysteresis loops for in-plane magnetization of the
heterostructures show signatures of an antiferromagnetic coupling
between the moments of the two LSMO layers in the superconducting
as well as the normal state of the spacer. The temperature
dependence of the exchange coupling energy shows a monotonic
growth followed by saturation on lowering the temperature. The
long range coupling was found to decrease exponentially with the
increasing thickness of the spacer layer. The suppression of
J$_{1}$ at T $<$ T$_{c}$ suggests inhibition of single electron
tunneling along the c-axis of YBCO as the in-plane superconducting
order parameter becomes non-zero.

\begin{acknowledgments}
This research has been supported by a grant from the Defense
Research and Development Organization, Government of India, and
the internal funding of I.I.T. Kanpur.
\end{acknowledgments}

%

\end{document}